# Modifying Hydrophilic Properties of Polyurethane Acryl Paint Substrates by Atomic Layer Deposition and Self-Assembled Monolayers

D. Beitner[a, b], I. Polishchuk[a], E. Asulin[b] and B. Pokroy[a*1]

a. Department of Materials Science and Engineering Technion - Israel Institute of Technology Haifa 32000, Israel.

b. Department of Materials Engineering, Ministry of Defense. Tel Aviv Israel.

A process of atomic layer deposition (ALD) combined with self-assembled monolayers (SAMs) was used to investigate the possible modification of polyurethane (PUR) paint surfaces. First, we used an ALD process to produce thin and uniform $Al_2O_3$ coatings of these surfaces at temperatures as low as 40 °C. We then successfully achieved the addition of 16-phosphono-hexadecanoic acid (16-PHA) SAMs to the $Al_2O_3$-coated paint samples. $Al_2O_3$ coatings reduced the contact angle of the PUR surfaces from 110 to 10°, accompanied by an initial hydrophobicity which however was not stable. over time. Addition of SAMs on the $Al_2O_3$ induced a sustained reduction in their contact angles to 60−70°, and aging of the samples revealed a further decrease to 25−40°. Testing of the $Al_2O_3$/16-PHA coating in a Weather-OMeter (WOM) revealed its durability even under harsh outdoor conditions. These experimental results show that by combining ALD with SAMs it is possible to produce durable coatings with modified hydrophilic/hydrophobic properties that are stable over time. The use of SAMs with different end-groups may allow fine-tuning of the coating's wetting properties.

**Key Words**

* Corresponding author at: Department of Materials Science and Engineering, Technion-Israel Institute of Technology, Haifa, Israel.
**Telephone: +972-4-8294584**
E-mail address: bpokroy@technion.ac.il

## 1. Introduction

Polyurethane (PUR) acryl polymer paints are widely used in the automotive and aerospace industries. There is growing interest in modifying the hydrophilic/hydrophobic properties of such paints for various practical applications. Most current research and development is focused mainly on various techniques for creating superhydrophobic coatings.[1,2] While some studies on the development of hydrophilic surface treatments have been reported, they were based mostly on surface structure and photo-induced mechanisms.[3] The development of a durable coating method for the creation of hydrophilic surfaces on polymer-based topcoats is important for adhesion and anti-fogging applications.

The difficulty of modifying the surface properties of PUR paint systems derives from the lack of reactive surface species in the paint substrate, the roughness of the paint surface, and the requirement that the coating not change the original color of the paint. Furthermore, any coating would have to be durable over time in outdoor conditions and maintain its hydrophilic properties when exposed to the elements.[4–7] In nature color on surfaces is achieved not only by pigmentation but also via structural coloring as well as ordered arrays of crystals such as Guanine crystals in fish scales.[8]

One way to create hydrophilic coatings is by depositing oxide layers on the substrate.[9–16] Since many oxides have hydrophilic surfaces, they make excellent candidates for hydrophilic coatings. However, most oxide deposition methods require high temperatures and are therefore not suited for polymer substrates.

Atomic layer deposition (ALD) is a two-stage vapor-deposition method that can be used to create uniform very thin coatings with high precision and control.[17–19] Since ALD processes can be performed at low temperatures, they allow oxides to be deposited on thermally sensitive substrates, such as polymers. Several works [12,14,16,20,21] have demonstrated oxide deposition on polymers using ALD. Oxides, however, owing to the instability of their surface species and their tendency to adsorb hydrophobic contamination, are of limited use as hydrophilic coatings for outdoor applications.

Stable surface modification may be achieved using self-assembled monolayers (SAMs). These monolayers are formed spontaneously on surfaces by adsorption and self-organization of organic molecules.[22,23] The wetting properties of SAMs can be controlled through the use of different functional groups, or fine-tuned by changing the length of the organic molecules.

SAMs are of limited use on polymer surfaces, however, because of the lack of surface chemistry needed for their adsorption. For SAMs to form on a polymer substrate, therefore, an intermediary layer with an appropriate surface species for SAMs formation is required. SAMs of alkylphosphonic acids have been shown to form dense, ordered and durable monolayers on oxide surfaces.[24–27] Accordingly, in our work we demonstrate the use of an intermediary oxide surface from ALD can be utilized for the formation of SAMs on polymer surface. This combined ALD / SAMs process can therefore be used to create thin and durable coatings for modifying the wetting properties of polymer substrates.

The goal of this work was to modify the wetting properties of PUR-based paint, through the use of ALD and SAMs, to create a durable hydrophilic coating.[12,28] The desired coating would reduce the contact angle of the paint to below 70° and maintain the reduction in contact angle under weathering, and not change the original color and hue of the paint.

## 2. Materials and Methods

### 2.1. Preparation of Paint

The paint used in this study was based on the commercial product Glasurit 923-155 MS-clear.[29] To improve the durability of the base varnish and ease its application, we mixed it with several additives: the surface-active material L77, the light-stabilizer Tinuvin®, Acematt® HK125 matting agent, quartz powder 30 µm, the anti-settling, anti-sagging agent BKY410, and Thinner #11.

### 2.2. Preparation of Paint Samples

Using a paintbrush, we applied 5 layers of paint on a rectangular (60 × 40 mm) silicon mold. The paint between each layer was allowed to partly harden for 30 minutes at 80 °C. We then filled the mold with an epoxy compound to create a backing for the sample. The samples were cured overnight at room temperature. They were then removed from the mold, washed with soap and water, and rinsed with deiononized water (DI). The samples were then post-cured at 120 °C in a vacuum oven at 1 mbar pressure for 48 hours.

### 2.3. Silicon Reference Samples

As control samples for some of the measurements, we used silicon wafers with the following parameters: native oxide surface, <100> surface direction, single side polished, 525 µm thickness, and P-type doping (Boron).

### 2.4. Atomic Layer Deposition Process

For ALD we used the PICOSUN® R-200 Standard ALD system. Trimethylaluminum (TMA; 97% (Sigma-Aldrich) and 98% (Strem Chemicals), both by Al content), was used with DI. The ALD processes were performed at a chamber pressure of ~1 Torr. Flow rates used were 150 SCCM for TMA and 200 SCCM for $H_2O$. Flow times of 'pulse / purge' (in seconds) used in the ALD process were 0.1 / 6.0 for TMA and 0.1 / 6.0 for $H_2O$. The different coating processes were performed over a temperature range of 50−120 °C.

### 2.5. Self-assembly Process

For the self-assembly process we used a 16-phosphono-hexadecanoic acid (16-PHA; 97%, Sigma-Aldrich) SAMs. All SAMs solutions were prepared in clean glass containers. For self-assembly on samples we used 2 procedures. In the first, samples (1 mM) were submerged in 99.9% pure ethanol for 24 hours, and were then thoroughly rinsed with DI water and ethanol. In the second procedure, samples (1 mM) were submerged for 24 hours in 99.9% pure ethanol with added DI water (1 wt%). Upon removal from solution the samples underwent thermal treatment at 80 °C for 1 hour. After the process they were thoroughly rinsed with DI water and ethanol.

### 2.6. Measurements of Wetting Properties

Contact angles were measured with a Theta Lite optical tensiometer (Biolin Scientific). A pipettor was used to place a 5-µl drop on the sample. To allow the drop to settle, measurements were taken 1 min after drop placement. Contact angles were measured at both sides of the drop. At least 15 drops on different areas of the sample surface were measured, and the results were averaged. Drop shape was analyzed on the basis of the Young-Laplace equation.

### 2.7. Measurements of Self-assembly Process

The formation of SAMs on ALD Alumina was measured using a quartz crystal microbalance with dissipation monitoring (QCM-D) Q-Sense® (Biloin Scientific). The QCM-D used a liquid flow cell model 401 (Biloin Scientific). The sensor crystal used was a ~ 5 MHz crystal with evaporated gold electrodes (Maxtec inc. CA USA). The original sensor frequency baselines were measured prior to surface preparation in an ethanol buffer solution, and the static contact angle of the sensor was measured by the Theta Lite.

The sensors surface was prepared by rinsing with DI water and ethanol several times. To create the alumina surface needed for the SAMs experiment a sensor holder was constructed. The holder was designed to hold the sensor, exposing only the gold upper surface, thus enabling coating of the sensor without damage to the electrode contacts. The sensor was coated with 100 pulses of alumina at 120 °C, using the same process used for paint samples. All solvents used were of analytical grade and were pumped using peristaltic pump (Reglo digital by Ismatec).

The experiment was performed as such:
1. An initial ethanol buffer solution was pumped into the cell for 30 minutes.
2. The buffer solution was replaced with a solution of 1mM 16-PHA in and pumped in a steady rate for 24 hours.
3. The SAMs solution was replaced with an ethanol buffer solution until frequency stabilized.

After the experiment, the sensors static contact angle was measured again.

## 2.8. Measurements of Color Properties

A Datacolor 245 photospectrometer was used to measure the color of samples. Measurements were performed at the 45°/0° geometry. The light source was a xenon flash lamp with a 400−700 nm wavelength range, using dual-beam mode. Reflected light from the sample was analyzed by a SP2000 spectral analyzer. The light spectrum was in the CIE1976 LAB color space system.[30]

Difference between the experimental and the reference measurement was calculated by the ΔE formula:

(1) $\quad \Delta E = \sqrt{L^2 + a^2 + b^2}$

According to the standard for measuring change in the color of paints over time, a change of ΔE > 3 is considered to be noticeable by the naked eye.

## 2.9. Measurements of Coated Surface Properties

Environmental Scanning Electron Microscopy (ESEM) micrographs were obtained using a Quanta 400 FEGSEM(FEI). To avoid charging of nonconductive samples, the micrographs were taken in low-vacuum mode at 90 Pa pressure. Because of the delicate and thin coatings studied, no conductive coatings were used in this work. Low pressure was established with DI water vapor to further reduce charging of samples. Images were obtained with a secondary electron detector. Energy-dispersive spectroscopies (EDS) were obtained using an EDAX liquid hydrogen-cooled sensor at 25 KeV beam current and 10 mm working distance.

Ellipsometry was performed using a Rudolph Ellipsometer AutoEL. Measurements were obtained from reference silicon wafers placed in the ALD reactor chamber with the samples.

## 2.10. Weathering of Coated Samples

Aging of samples was accelerated using a Weather-Ometer (WOM) Ci-5000 (Atlas). A xenon arc lamp system equipped with an inner and outer Type S borosilicate filter was used to simulate the solar spectrum.

We used a 2-part cycle program. In the first part, the lamp was kept operating at a total output of 0.37 W/m^2 for 102 minutes, in a chamber maintained at 50 °C and relative humidity (RH) of 55%. In the second part the lamp was off, and an active water spray was used to keep the chamber at 50 °C and 95% RH. The weathering program was derived from the ASTM D6695 standard for automotive paint systems.[31]

## 3. Results and Discussion

### 3.1. Growth rate of ALD alumina at low temperatures

To measure the growth rate of ALD as a function of temperature, we used silicon wafers as reference samples. $Al_2O_3$ coating by ALD was performed for 600 cycles at 100 °C, 110 °C and 120

°C. At several points for each sample, coating thickness was measured using an infrared ellipsometer. Samples coated at 100 °C showed the smallest variation in coating thickness. The measured growth per cycle (GPC) for $Al_2O_3$ on the reference samples is shown in Table I. The measured GPC is close to the reported growth rate of 1.1−1.2 Å/cycle.[32,33]

*Table I. Growth rate of ALD alumina on silicon wafers measured over 600 cycles at different temperatures.*

| Temperature (°C) | 100 | 110 | 120 |
|---|---|---|---|
| GPC (Å/cycle) | 0.99 | 1.06 | 1.3 |

### 3.2. Hydrophilic properties of alumina coatings

PUR paint samples were coated with 600 cycles of $Al_2O_3$ at 100 °C. Contact angles were measured before the coating process and were monitored for 24 hours after it. When measured immediately after coating followed by full wetting, the $Al_2O_3$ samples exhibited super hydrophilic properties. Contact angles of the coated samples were found to increase with time after returning to pre-coating (hydrophobic) values after 24 hours, as shown in Figure *1*.

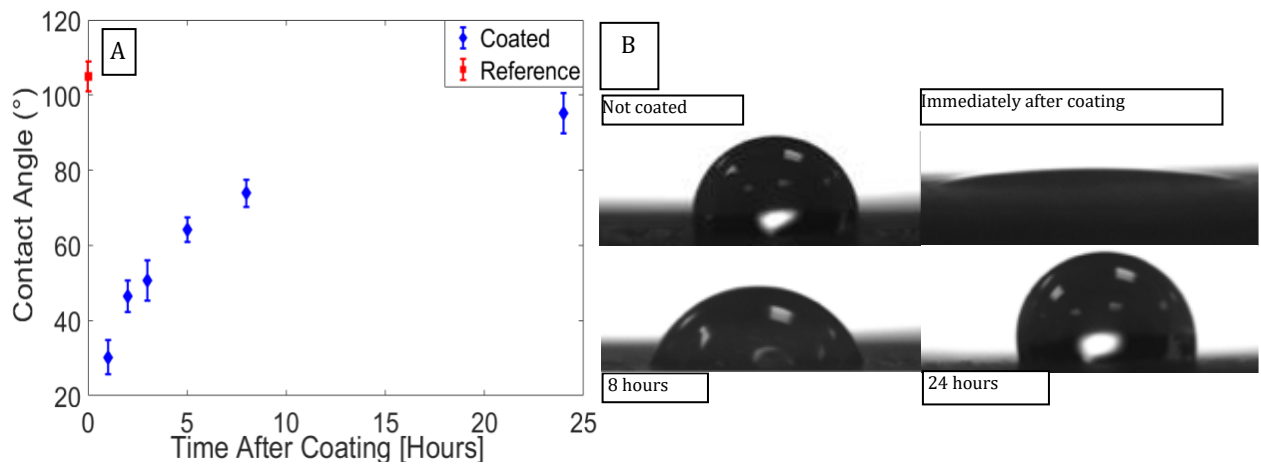

*Figure* 1*. PUR sample coated with 600 cycles of alumina at 100 °C. (A) Static contact angle of DI water was measured over time using the Theta device. Immediately after coating, the sample*

*exhibited total wetting behavior and the contact angle was too low to measure accurately. (B) Images taken with Theta Lite at different times are recorded. Evidence for degradation of the hydrophilic property of the coating can be seen over time.*

The silicon reference samples exhibited similar wetting-transition behavior when coated with $Al_2O_3$ using ALD (see Additional Information), indicating that contamination of the coating surface by the substrate is not the cause of the increase in contact angle observed for the samples. An increase in contact angle with time after ALD of $Al_2O_3$ was previously reported for polydimethylsiloxane coated with $Al_2O_3$.[34]

The $Al_2O_3$ coating was observed to lose its hydrophilic properties over time in ambient conditions. In order to test the effect of adsorption of contamination to the $Al_2O_3$ surface on its hydrophilic properties the contact angle measured over time. PUR paint samples were coated and placed in vacuum. The sample's contact angle was measured over time and compared to that of the samples kept under ambient lab conditions. The increase in contact angle observed over time for both types of samples was similar (Figure 2). This indicates that the change in the contact angle does not stem from the adsorption of hydrophobic contamination to the surface.

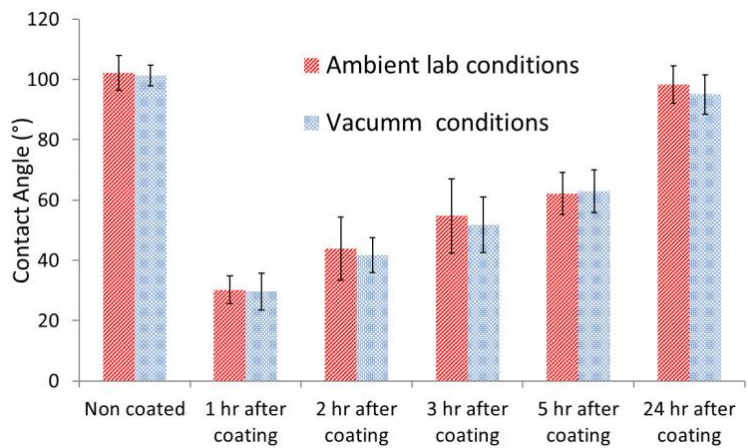

*Figure 2. PUR paint samples were ALD coated by 600 cycles of $Al_2O_3$. One sample was kept in ambient lab conditions and the other in a vacuum chamber. Both samples show identical rate of loss of hydrophilic properties over time.*

The increase of the contact angle over time indicates an increase of the surface energy [35]. A possible explanation is the shift of surface polar groups into the bulk of the $Al_2O_3$ in order to reduce the surface energy of the alumina.[36–38] In order to create a long lasting hydrophilic surface a more stable surface species is required.

### 3.3. ESEM images of ALD alumina on PUR paint

ESEM images of the coated samples were obtained before and after the coating. Small cracks (~550 nm) were observed in the alumina layer of samples, as seen in Figure 3 for a sample coated with 600 cycles of alumina at 80 °C.

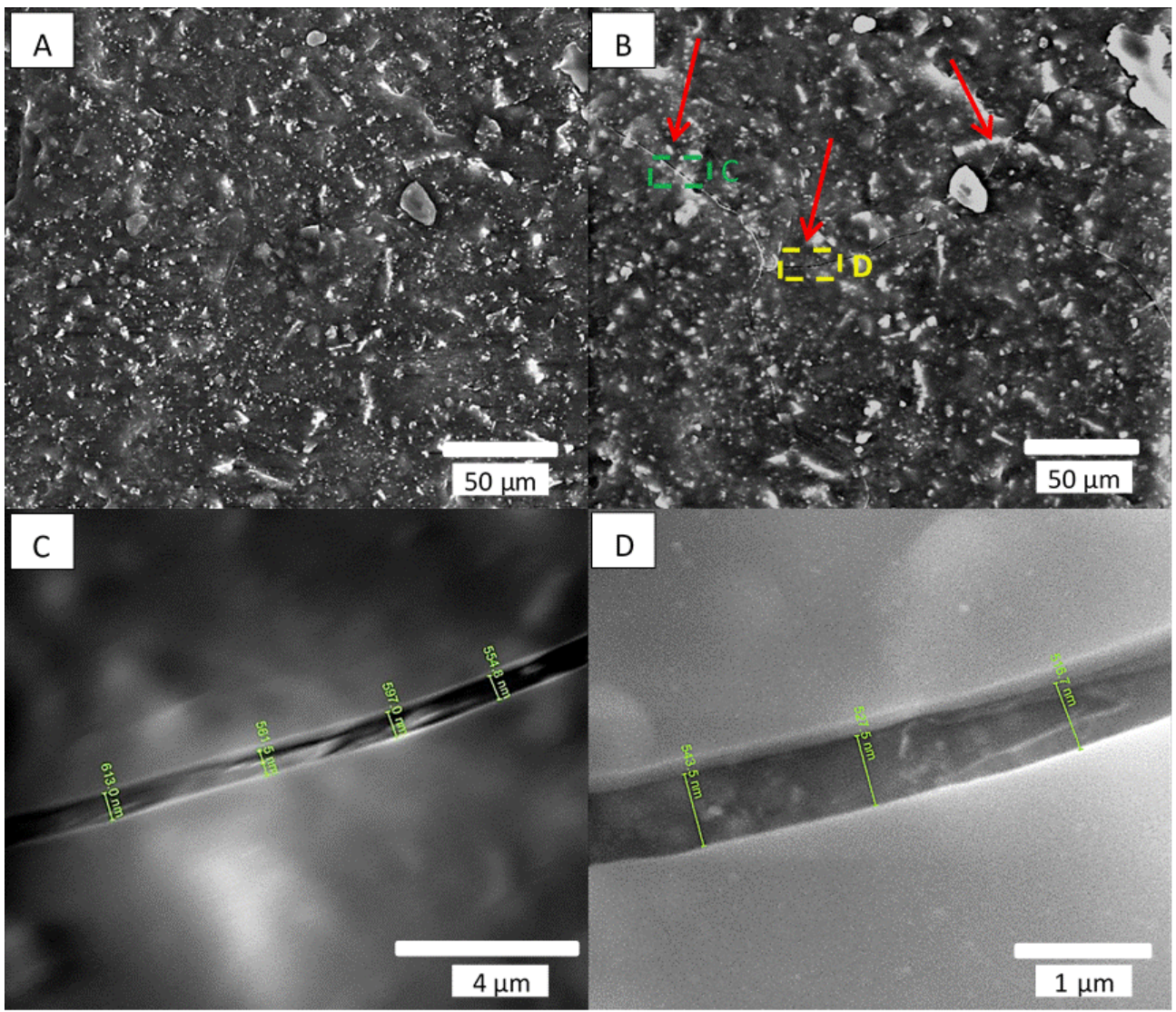

*Figure 3. Scanning electron micrograph (SEM) secondary electron images of a paint sample coated with 600 pulses of alumina at 80 °C. (A) Sample before coating. (B) Sample after coating; red arrows show cracks in the coating. (C) Magnified image of the yellow rectangle in B, showing*

*a crack of ~600 nm. (D) Magnified image of green rectangle in B, showing a crack of ~530 nm in the alumina. The paint surface can be seen through the crack.*

EDS measurement confirmed that the layer shown in Figure 3 C-D is $Al_2O_3$, (see supplemental information Figure S 1). The coated sample initially showed hydrophilic properties, which—as observed with other coated samples—reverted to uncoated values after 1 week, as seen in **Error! Reference source not found.**.

*Table II. Static contact angle measurements on a sample coated with 600 pulses of alumina at 80 °C. Measurements were taken before coating, immediately after coating, and 1 week after coating. The sample was measured with a 1-µl and a 5-µl drop of DI water applied with a pipettor.*

| Drop size | Uncoated | Immediately after coating | 1 week after coating |
|---|---|---|---|
| 1 µL | 111 ± 4 ° | 37 ± 7 ° | 102 ± 6 ° |
| 5 µL | 111 ± 4 ° | 39 ± 9 ° | 100 ± 6 ° |

Cracks in the alumina coating were suspected to be a possible cause of loss of the hydrophilic property of the coatings over time. We therefore compared the cracks seen in the alumina coating on a sample immediately after coating and 1 week after coating, as seen in Figure 4.

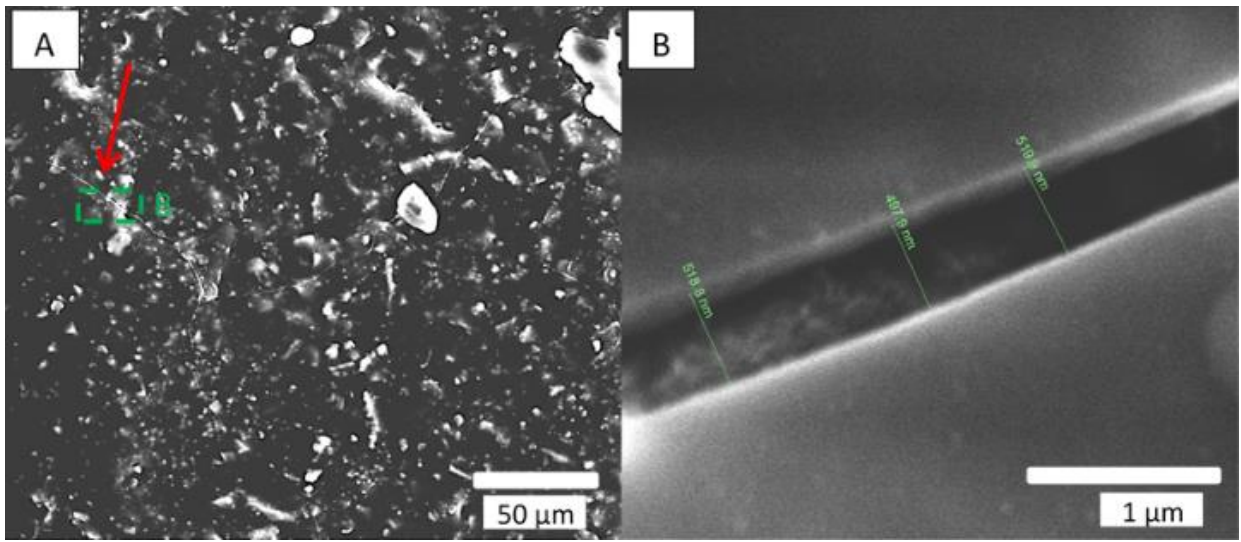

*Figure 4. SEM secondary electron micrograph of paint sample coated with 600 cycles of alumina at 80 °C. (A) Secondary electron micrograph of the crack in the alumina coating 1 week after the coating was applied. (B) Magnified image of the crack seen in the green square area.*

*Compared to the same measurement performed immediately after coating (see Figure 2B), the crack shown here has not changed significantly in thickness.*

Kemell et al.12 and Spagnola et al.34 reported similar cracks for the ethylene-tetrafluoroethylene copolymer coated at 80 °C with 300 cycles and for polydimethylsiloxane coated at 25 °C with 100 cycles. Kemell et al. attributed such cracks to handling damage, while Spagnola et al. attributed them to thermal effects during deposition or to swelling of the latter polymer due to precursor adsorption. Alternatively, the observed cracking might be explained in terms of a thermal mismatch between the alumina and the paint substrate. To test this possibility, we coated a sample at a lower temperature and examined it for cracks. A decrease in coating temperature upon their cooling to room temperature should reduce the thermal stress produced between the coating and the paint and should thus result in fewer and smaller cracks in the sample. Immediately after coating of our paint sample with 600 cycles of alumina at 40 °C, hydrophilic properties of total wetting were revealed by the Theta device on certain areas, which returned to the original wetting behavior, with a measured contact angle of ~ 100 °after 24 hours. As seen in **Error! Reference source not found.**, the sample coated at 40 °C showed less cracking of the coating, and (as seen in Figure 5) the measured cracks were smaller (~93 nm) than those observed in the samples coated at 80 °C (~564 nm).

*Table III. Average crack size measured in paint samples coated with 600 pulses of alumina at 80 °C and 40 °C.*

| Coating temperature (°C) | Crack size (nm) |
|---|---|
| 80 | 560±39 |
| 40 | 90±13 |

This result indicates that thermal mismatch between the layers is probably the main cause of the observed cracking.

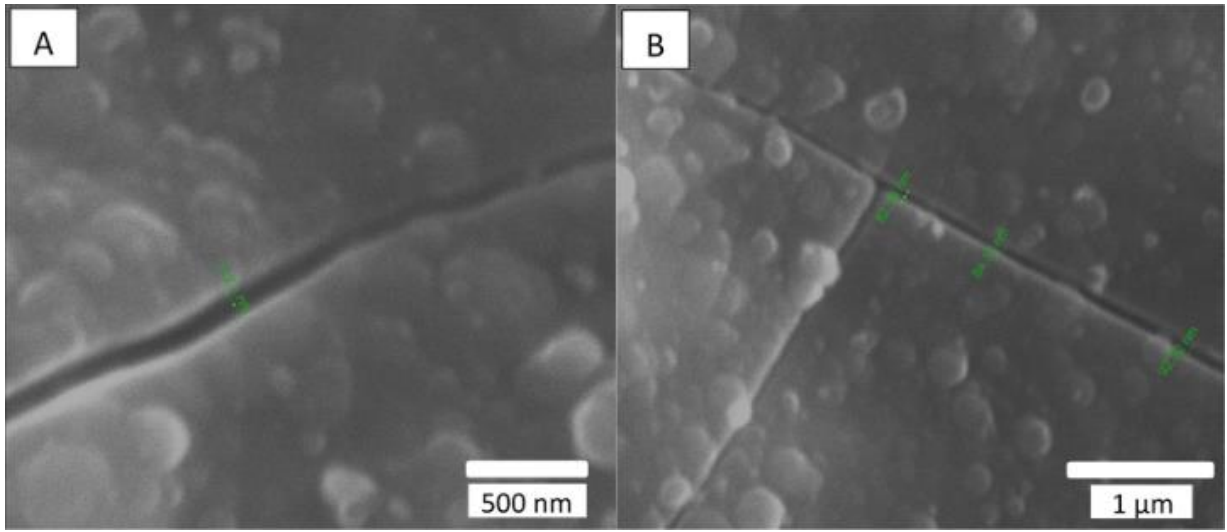

*Figure 5. SEM secondary electron micrograph of the surface of a paint sample coated with 600 cycles of alumina at 40 °C. Both (A) and (B) show cracks of about 90 nm in width in the alumina layer.*

The cracks in the alumina layer deposited on paint samples did not seem to develop further with time after coating, and the samples coated at lower temperatures seemed to demonstrate less cracking and smaller cracks, but they still lost their hydrophilic properties with time. This suggested that cracking of the alumina coating observed in coated samples over time under ambient conditions is probably not the reason for the increase in contact angle.

The results of the first part of this work showed, therefore, that although an ALD procedure can be used to successfully deposit a thin uniform layers of aluminum oxide on PUR paint substrates, the hydrophilic properties of the resulting layers are not stable under ambient conditions over time. This means that the ALD process by itself is not sufficient to create a hydrophilic coating for our paint substrate, but that the initial high degree of hydrophilic behavior exhibited by the film could possibly be used as a basis for the formation of other surface-modifying processes, such as SAMs growth.

### 3.4. SAMs Growth on ALD of Alumina

To prepare paint samples for SAMs processing we used a smooth silicon mold with polymer backing. Samples were coated with 100 cycles of alumina at 100 °C and exhibited initial hydrophilic behavior (complete wetting) after coating. Immediately after coating, the samples were placed in the container together with 1 mM 16-PHA SAMs solution in ethanol. After different time periods in the SAMs solution, starting at 10 minutes, a sample was removed, washed with DI water and ethanol, and the contact angle was measured. The effect of SAMs growth time on the wetting angle is shown in Figure 6.

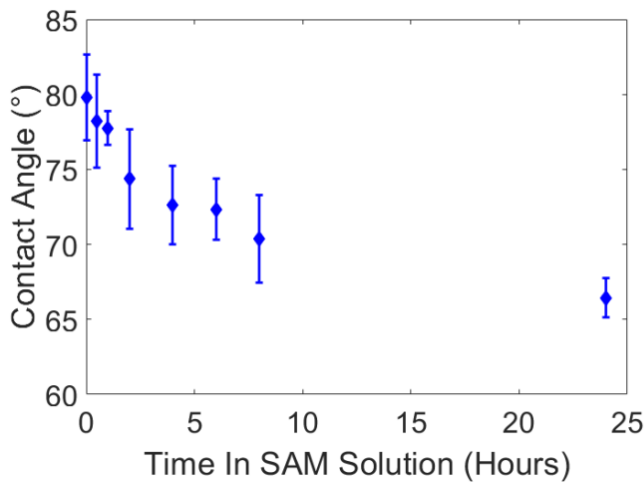

*Figure 6. Smooth paint samples were coated with 100 pulses of alumina at 120 °C. The samples were placed in a SAMs solution of 1 mM 16-PHA in ethanol. At specified time periods a sample was removed from the solution, washed with DI water and ethanol, and its contact angle was measured. Contact angles of the samples decreased with time in the SAMs solution.*

Initially the contact angles of samples were about 80°. After 24 hours in the SAMs solution this decreased to ~66.5°, indicating the formation of a hydrophilic SAMs layer on the alumina surface in the SAMs/ethanol solution over time.

To test the durability of the SAMs under ambient conditions, samples were coated with ALD alumina (100 cycles at 100 °C) and immediately treated with SAMs (by immersion in 1 mM 16-PHA solution in ethanol for 24 hours followed b**y** thermal treatment for1 hour at 80 °C). Samples' contact angles were measured immediately after the SAMs growth and again after1 month under ambient conditions.

As shown in **Error! Reference source not found.**, samples coated with ALD alumina and then treated with SAMs maintained their reduced contact angle after a month in ambient conditions. This indicates a more stable hydrophilic surface than that produced by coating with ALD without SAMs treatment.

*Table IV. Contact angle measurements in samples coated with ALD alumina for 100 cycles at 100 °C and then subjected to SAMs treatment by immersion for 24 hours in 1mM 16-PHA ethanol solution followed by thermal treatment for 1 hour at 80 °C. Contact angles were measured before and after coating and again after a month under ambient conditions.*

|  | 100 cycles at 100 °C with SAMs | Uncoated | Coated | 1 month in ambient conditions |
|---|---|---|---|---|
| Contact angle (°) | 110±5 | 62±5 | 65±4 |

### 3.5. SAMs Growth Kinetics

Kinetics of 16-PHA SAMs growth on ALD alumina was measured by a QCM-D experiment. The experiment measured the change in the resonance frequency and dissipation factor of the sensor over a 24-hour growth period of 16-PHA of SAMs. Graphs of change in the resonance frequency and dissipation factor for the third resonance mode can be seen in the supplementary data (Figure S 2).

The Sauerbrey equation correlates the change in resonance frequency to change in the sensors mass for a stiff oscillating resonator (very small values of dissipation compared with frequency change $\frac{\Delta D}{\Delta f}$) [39,40]:

$$\Delta f = -\frac{f}{t_q \rho_q} \Delta m = -n \frac{2 f_0^2}{v_q \rho_q} \Delta m = -n \frac{1}{C} \Delta m \quad (1)$$

Where $v_q$ is the wave velocity (speed of sound) in the quartz, $t_q$ is the thickness of the quartz sensor, $\rho_q$ is the quartz density and n is the resonance overtone (n=1 fundamental resonant frequency, n=3 third overtone and so on).

After ALD deposition of alumina on the sensors surface, the resonance frequency of the sensor changed by about ~-229. The change in dissipation after the ALD alumina coating was about ΔD≈0.11x10⁻⁶, which is much smaller than the observed change in frequency

$$(\frac{\Delta D}{\Delta f} << 0.1 \cdot 10^{-6} Hz^{-1}).$$

The mass of ALD alumina deposited on the sensor was calculated from the frequency using the sauerbrey equation and the parameters for the quartz sensor. for the ALD alumina. The results were compared to results obtained from IR ellipsometer measurements performed on a silicon wafer coated under the same conditions, as seen in **Error! Reference source not found.**.

*Table V. comparison of QCM data for sensor coated with 100 cycles of ALD alumina at 120 °C to data obtained from IR ellipsometer measurement of silicon wafer coated under the same condition. The density of alumina was taken as 3.0 gr/cm3 in order to calculate thickness from QCM data.*

|  | Thickness (nm) |
| --- | --- |
| Ellipsometry measurement | 15.6±0.5 |
| Calculated from QCM data | 12.84±0.01[a] |

[a] The thickness was calculated using a density of 3.0 gr/cm3 for the alumina. [32]

The thickness calculated for the QCM data is lower than that measured for the silicon wafer. This could be a result of the Sauerbrey equation underestimating the deposited mass or actual difference in the coating thickness between the samples. Difference in deposited thickness between the samples could be a result of surface energy difference between the gold and the silicon wafer. Furthermore, the mask which was used to enable selective coating of the sensor in the ALD might have obstructed the flow of the precursor, thus lowering the growth rate per a cycle achieved for the sensor compared with the silicon wafer.

The baseline frequency of the alumina coated sensor was measured after ethanol buffer solution flow. There was an increase in the measured resonance frequency indicating a mass decrease on the sensor. The loss of mass during the buffer baseline measurement could be a result of the buffer solution removing contamination and residual reactants from the coating surface.

A stable frequency baseline in the buffer solution was reached, after which, a solution of 1 mM 16 PHA in ethanol was introduced to the flow cell. The resonance frequency of the sensor drops rapidly after the introduction of the 16-PHA SAMs into the flow cell, as seen in **Error! Reference source not found.**, indicating an initial rapid mass increase. This may be the result of 16-PHA molecules adsorbing on the alumina surface.

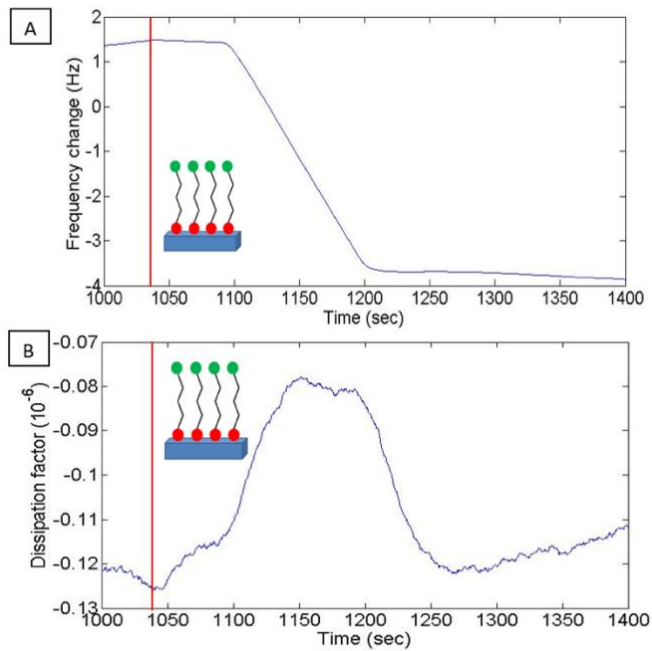

*Figure 7. QCM results for 16-PHA SAMs growth on ALD alumina experiment. Results for 1000-1400 seconds showing the introduction of 16-PHA SAM into the flow cell. The red line in both graphs indicates the point at which the buffer was switched to 16-PHA SAMs solution. (A) Change in resonance frequency of resonance mode 3 showing a rapid drop in frequency after switch to SAMs solution. (B) Change dissipation factor of resonance mode 3 there is a small peak in the dissipation probably due to initial dampening during adsorption of SAMs molecules to the sensor.*

The small peak seen in the dissipation (Figure 7B) is probably due to initial dampening due to the increase of the viscosity of the surface liquid on the sensor, during the adsorption of the SAMs to the sensors surface.

After the initial rapid drop in resonate frequency, there is a more moderate decrease in frequency over time, indicating a slower rate of mass increase on the sensor.

The diagram of the process as seen in

Figure 8 shows two distinct phases during the formation of the SAMs. An initial low viscosity phase and a higher viscosity phase. This fits the described process for SAMs formation at a temperature under the triple point where process starts with a low densities SAMs phase which transitions through a mixed low-high density coexisting region.25,41

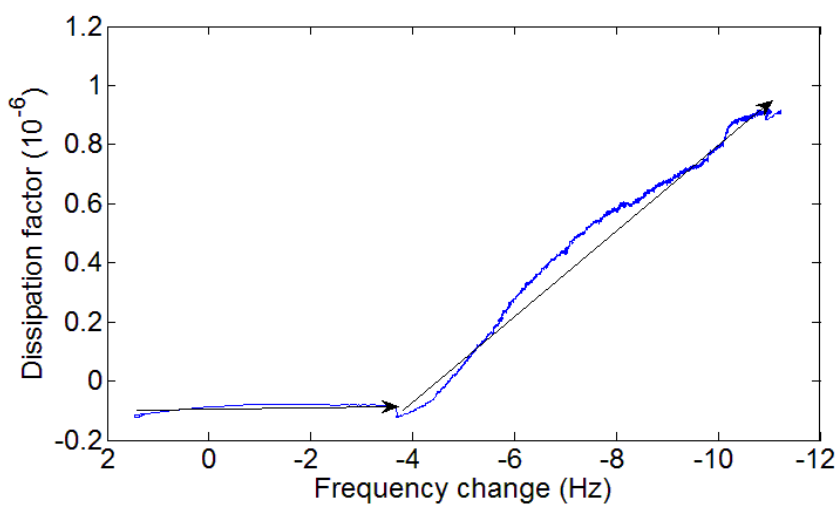

*Figure 8. diagram of SAMs formation process on ALD Al2O3. The black arrows mark the flow of the process with time. Two distinct phases can be seen in the diagram an initial low viscosity phase marked by a very low increase of dissipation with decrease of frequency (increase of mass) and a higher viscosity phase with a noticeable increase of dissipation with frequency.*

After 24 hours the solution was changed back to ethanol buffer solution as seen Figure S 3. A small increase in frequency was observed after switching back to buffer solution, indicating a mass decrease, possibly due to SAMs Bilayers being removed by the buffer solution.

The experiment was stopped after resonance frequency stabilized for the sensor. Due to the relatively small change in resonance frequency ($\frac{\Delta f}{f_0} < 0.02$) and the low dissipation factor ($\frac{\Delta D}{\Delta f} < 0.1 \cdot 10^{-6} Hz^{-1}$), the use of Sauerbrey equation was assumed to be valid for calculating the change in the mass of the sensor after SAMs formation. The measured increase in mass due to SAMs formation on the sensor was calculated as 247±1 ng.

The theoretical maximum surface density (maximum number of SAMs molecules per a square nanometer of surface area) for 16-PHA SAMs on alumina is $\sigma_{max} = 4.35 \frac{molecule}{nm^2}$ .[42,43]

Using the known dimensions of the sensor (12 mm diameter) and the molar mass of the 16-PHA molecule ($Mw_{16-PHA} = 336.4 \frac{gr}{mole}$) we can calculate the maximum theoretical mass for deposited 16-PHA SAMs:

$$(2) \quad m_{T-max} = \frac{\sigma_{max} \cdot A_{sensor} \cdot Mw_{16-PHA}}{N_A}$$

Where $A_{sensor} = 1.13 \cdot 10^{14} nm^2$ is the area of the sensor and $N_A$ is Avogadro's constant. Comparing the theoretical maximum mass to that calculated from the QCM data we get that the cover ratio is in the order of ~0.9. The cover ratio calculated form the QCM data matches other reported works [43,44], this indicates that the 16-PHA SAMs creates a closely packed monolayer on the alumina surface after an immersion time of 24 hours in SAMs/ethanol solution.

### 3.6. Weathering of Coated Paint Sample

To test the durability of 16-PHA/SAMs on alumina and the ability of this combination to coating to maintain its hydrophilic properties over time, we placed samples in a WOM. **Error! Reference source not found.** shows the different sets of samples prepared for weathering.

*Table VI. Set of samples prepared for WOM testing of the durability of the coating process. Samples were prepared with several different thicknesses of alumina coating and were then immediately placed in SAMs/ethanol solution for 24 hours.*

| Samples | Coating temperature (°C) | Thermal treatment |
|---|---|---|
| Group | - | - |
| 1 | 80 | - |
| 2 | 100 | - |
| 3 | 100 | 1 hour at 80 °C |

Sample contact angles and hues were measured immediately after SAMs growth and after weathering in WOM. Contact angle results were compared to a painted reference sample that was placed in the WOM with no coating.

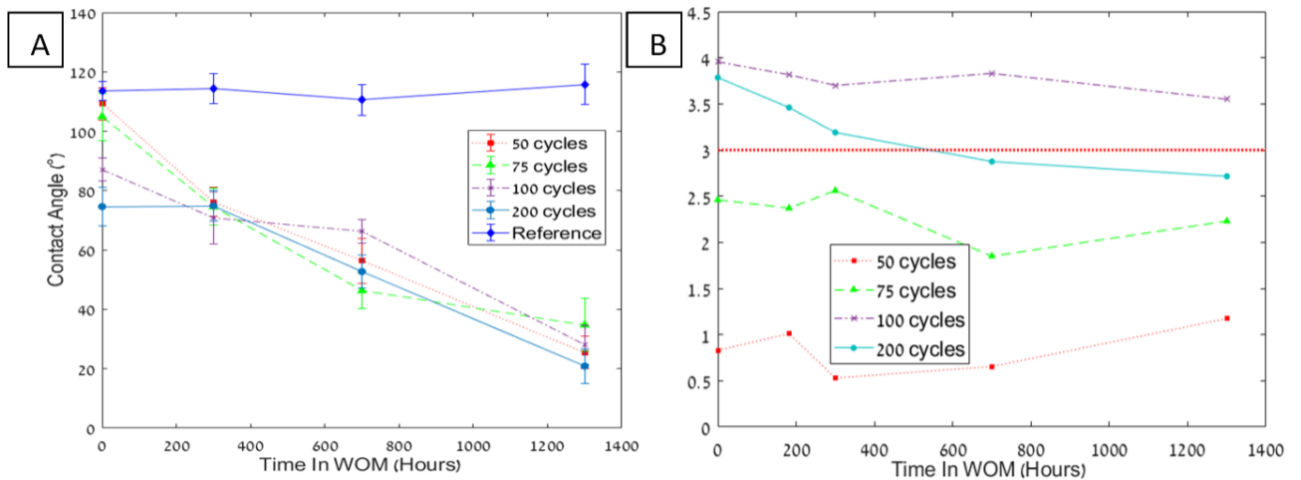

*Figure 9. Results for sample group 1 coated at 80 °c with no thermal treatment. Measurements of contact angle (A) and hue (B) are shown.*

Figure 9 presents the results in group 1 (**Error! Reference source not found.**) for samples coated at 80 °C, showing that the initial contact angle in the sample decreased with increasing thickness of the ALD coating measured in cycles**.** Group 1 samples coated with 50 cycles showed no decrease in contact angle immediately after being coating with ALD alumina, and only a minor decrease in contact angle after SAMs formation (Figure 9A). This might indicate that, irrespective of SAMs, a uniform coating is fully formed on the paint only after more than 100 cycles of ALD alumina at 80 °C.

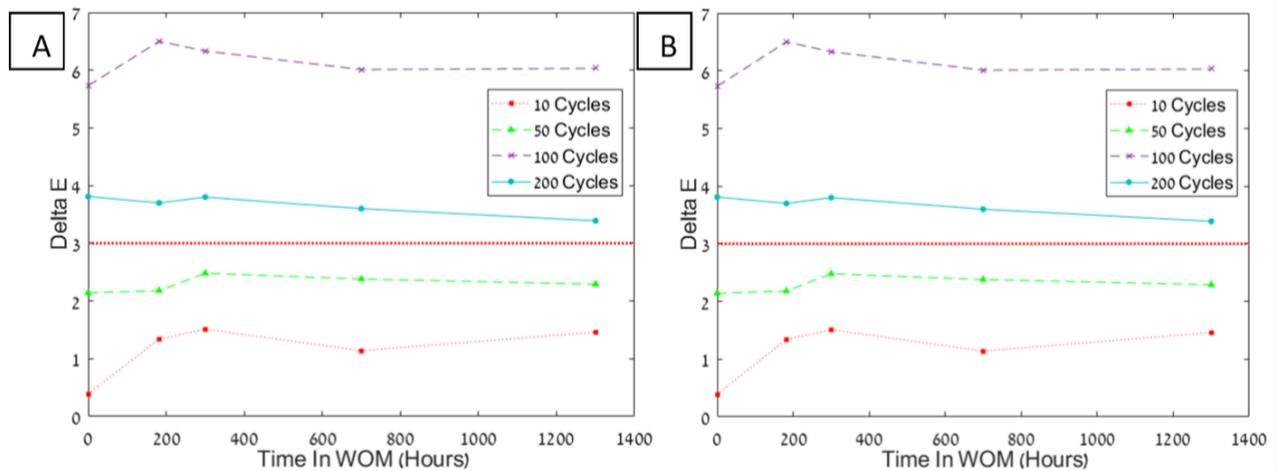

*Figure 10. Results for sample group 2 coated at 100 °C with no thermal treatment. Measurements of contact angle (A) and hue (B) are shown.*

As can be seen in the contact angle results for samples with SAMs formation (Figure 11), the contact angle of the sample decreases with weathering over time. This was observed for all samples, and may be a result of the SAMs forming a more ordered phase over time due to heating in the WOM. The more ordered the SAMs phase, the more functional group it contains on the monolayer surface, thus increasing the surface energy and reducing the contact angle. Another possible explanation is that hydrophilic salts from the water are adsorbed to the sample surface during weathering.

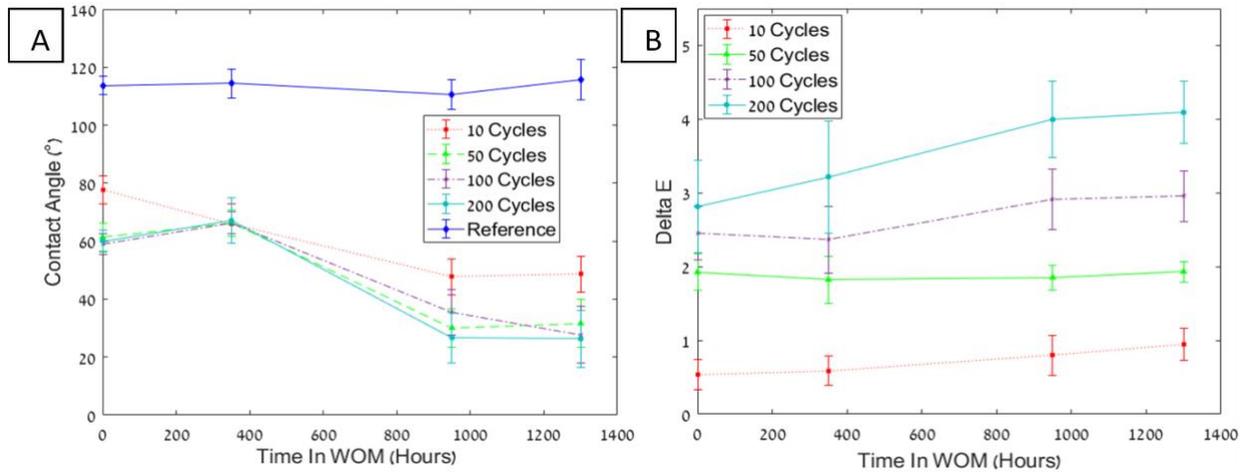

*Figure 11. Results for sample group 3 coated at 100 °C with thermal treatment. Measurements of contact angle (A) and hue (B) are shown.*

Figure 11 depicts the results in group 2 (Table 5) for samples coated at 100 °C, and shows that for ALD alumina coating at 100 °C, an effect of the number of ALD cycles is seen only for fewer than 50 cycles of coating thickness.

The results for group 3 (Table 6) of samples coated at 100 °C with thermal treatment, as seen in Figure 11, show a lower initial contact angle for samples after SAMs formation. The lower contact angle might reflect a more ordered SAMs phase due to the thermal treatment.

The initial color change of the samples increases with coating thickness. For all sample groups, a sample coated with 100 or 200 pulses has a ΔE >3, which exceeds allowed values.

Samples ALD-coated with alumina and then exposed to 16-PHA SAMs formation process maintained hydrophilic properties even after 1000 hours of weathering. Sample roughness was measured using a confocal microscope on a 500x500 $\mu m^2$ area before and after weathering, Sample initial averaged roughness is high Sa=10.3±2.0 μm due to the innate roughness of the paint surface. There was no significant change observed in surface roughness after the weathering process.

## 4. Conclusions

A durable hydrophilic coating for PUR paint substrates was successfully developed and demonstrated. This was achieved by the use of advanced coating techniques employing ALD and SAMs growth.

In this work, successful coating of a PUR paint substrate with a high degree of roughness was achieved using a low-temperature aluminum oxide ALD process with TMA/water precursors. The initially hydrophilic behavior of the resulting oxide layer diminished rapidly over time, possibly owing to contamination of the oxide's surface or a shift of surface polar groups into the bulk of the aluminum oxide.

The formation of 16-PHA SAMs on paint samples with aluminum oxide ALD was shown here to create a stable hydrophilic layer under ambient conditions. Investigation of the growth kinetics of the SAMs showed a time-related decrease in the contact angle for paint coated with alumina in the SAMs solution, further supporting the suggestion that SAMs had formed on the sample.

The combined ALD/SAMs coating process produced a large reduction in the contact angles of our paint samples from an initial 110 to 65°. SAMs growth was observed on several different thicknesses of alumina coatings on the paint, and was shown to maintain its hydrophilic properties over time, and for more than 1000 hours of weathering. The contact angle was also shown to improve with time in weathering, reaching ~20−30 °. This might point to continued formation of a SAMs-ordered phase during weathering, or adsorption of polar contamination on the sample's surface.

Overall, the method shows promise for use in the production of durable hydrophilic coatings on PUR paint. The coatings created here were shown to be suitable for outdoor applications. Furthermore, the coatings were shown to be thin enough not to change the initial color of the paint when applied (ΔE <3). This multilayer coating process can be further extended by the use of alkylphosphonic acid SAMs with other functional groups, enabling manufacturers to customize the PUR paint surface for a variety of different applications.

## 5. Acknowledgements

We thank the Technion-Israel Institute of Technology for the support in performing this research.

## Supplementary Information

1. EDS results

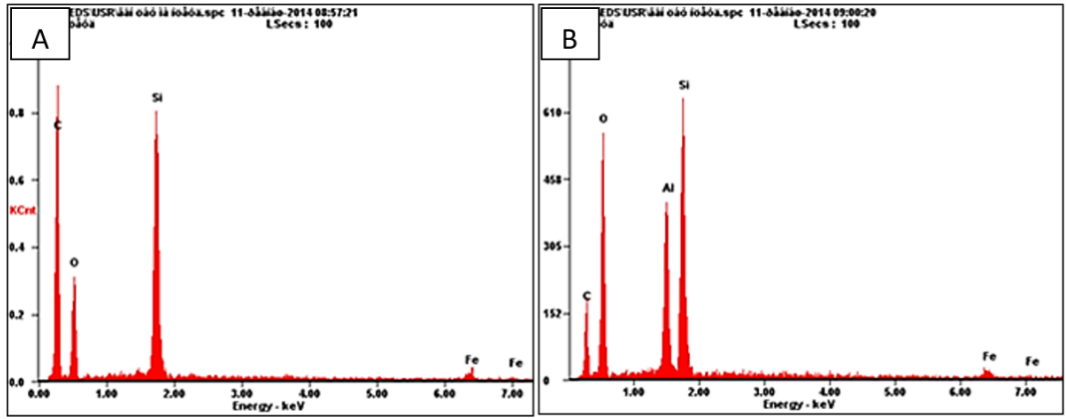

*Figure S 1 - In (A) and (B), EDS measurements are shown for the area of Figure 2A and 2B, respectively. The measurements show an increase in aluminum and oxygen after coating, as expected for the alumina layer.*

2. QCM results

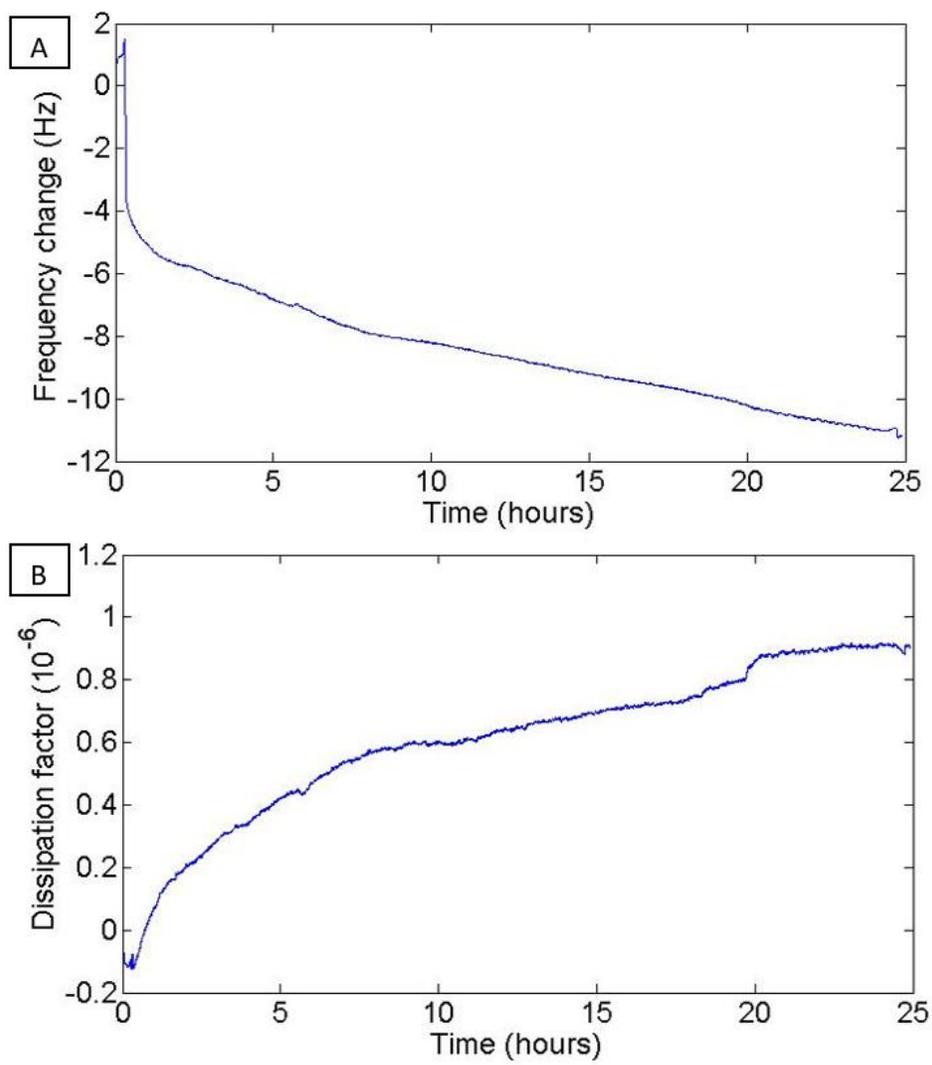

*Figure S 2 - QCM results for 16-PHA SAM growth on ALD alumina experiment. Data was filtered with a moving average of 300 data points. (A) Change in resonance frequency of resonance mode 3. (B) Change dissipation factor of resonance mode 3.*

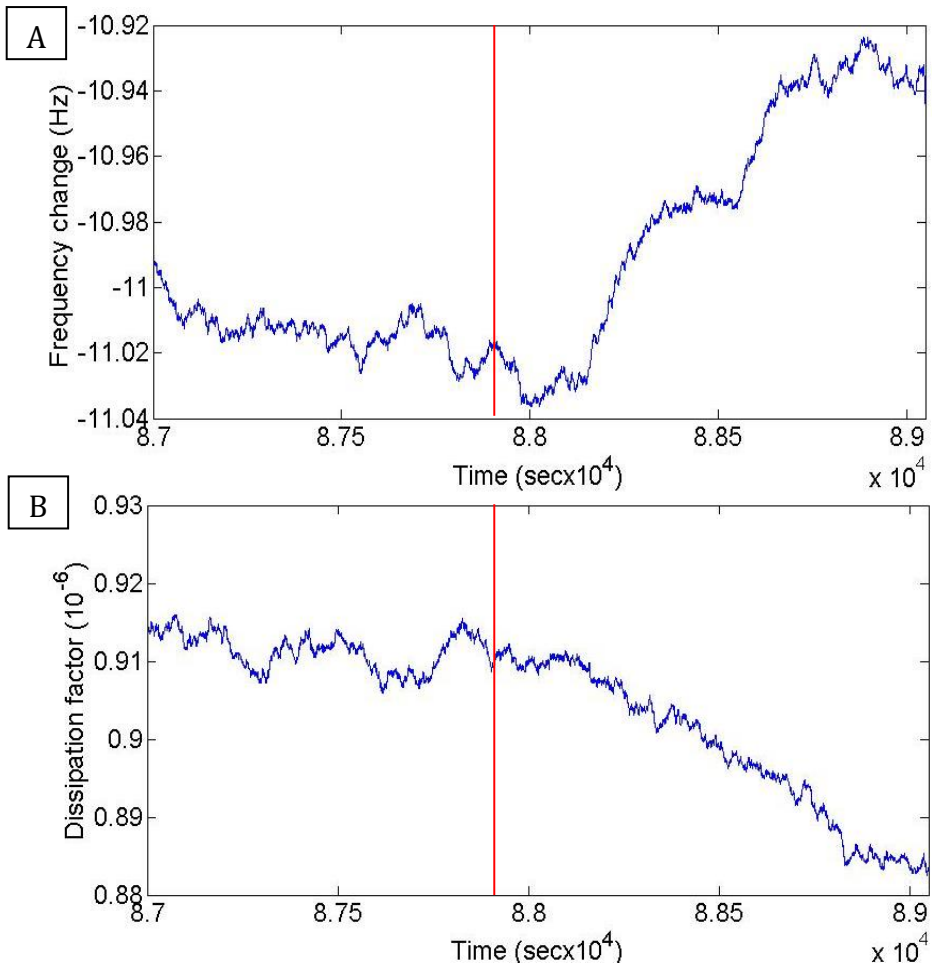

*Figure S 3 - QCM results for 16-PHA SAM growth on ALD alumina experiment. Results for 87000 89000 seconds showing the switch back to buffer solution after the 24 hour period. The red line in both graphs indicates the point at which the 16-PHA SAM solution was switched to buffer solution. (A) Change in resonance frequency of resonance mode 3 showing a small increase in frequency after switch to buffer solution. (B) Change dissipation factor of resonance mode 3.*